# Electrical tuning of layer-polarized exciton gases in WSe$_2$ bilayers

Zefang Wang, Yi-Hsin Chiu, Kevin Honz, Kin Fai Mak*, and Jie Shan*

Department of Physics and Center for 2-Dimensional and Layered Materials
The Pennsylvania State University, University Park, Pennsylvania 16802-6300, USA

*Correspondence to: kzm11@psu.edu; jus59@psu.edu

**Abstract**

Van der Waals heterostructures formed by stacking two-dimensional atomic crystals are a unique platform for exploring new phenomena and functionalities. Interlayer excitons, bound states of spatially separated electron-hole pairs in van der Waals heterostructures, have demonstrated potential for rich valley physics and optoelectronics applications, and been proposed to facilitate high-temperature superfluidity. Here, we demonstrate highly tunable layer-polarized excitons by an out-of-plane electric field in homo-bilayers of transition metal dichalcogenides. Continuous tuning from negative to positive layer polarity has been achieved, which is not possible in hetero-bilayers due to the presence of large built-in interfacial electric fields. A large linear field-induced redshift up to ~ 100 meV has been observed in the exciton resonance energy. The Stark effect is accompanied by an enhancement of the exciton recombination lifetime by more than two orders of magnitude to > 20 ns. The long recombination lifetime has allowed the creation of a layer-polarized exciton gas with density as large as $1.2 \times 10^{11}$ cm$^{-2}$ by moderate continuous-wave optical pumping. Our results have paved the way for realization of degenerate exciton gases in atomically thin semiconductors.

**Keywords:** 2D semiconductors, van der Waals' heterostructures, layer-polarized excitons, quantum confined Stark effect

Stacking two-dimensional (2D) atomic crystals into van der Waals heterostructures has allowed the design of fascinating new properties while maintaining many of the unique features of the constituent layers [1,2]. Semiconductor heterostructures such as transition metal dichalcogenide (TMD) bilayers are particularly interesting for photonics and optoelectronics because of the strong excitonic effect and valley-dependent physics [3-8]. Ultrafast charge transfer [9,10] and formation of interlayer excitons [7,11] have been demonstrated in photoexcited TMD heterostructures. In particular, the formation of interlayer excitons has enabled a much longer exciton lifetime [7,11] and valley depolarization lifetime [7,12], holding promise for valley optoelectronics [5,6,13,14]. So far the studies have focused on hetero-bilayers [6,7,9-13,15,16], such as $MoSe_2/WSe_2$ and $MoS_2/WS_2$, that are formed by two distinct TMDs. The large built-in interfacial electric field in these systems [6,7,9-13,15,16] has, however, significantly limited the electrical tunability of the interlayer excitons. On the other hand, the built-in electric field is absent in pristine homo-bilayers that are formed by two identical TMD monolayers. The weak interlayer interaction further allows the separation of the bilayer systems into monolayer-like components by a modest vertical electric field [17-21]. While earlier experimental studies have observed insignificant field effects on intralayer excitons [18,19,21], the application of a vertical electric field in our high-quality bilayer samples has allowed the generation of layer-polarized exciton gases (Fig. 1a) with widely tunable properties including the exciton dipole orientation, emission energy and lifetime, opening up new opportunities for tunable photonic and optoelectronic devices [5,6,13,14] and creation of high-density exciton gases [22].

We fabricate dual-gate field-effect devices of $WSe_2$ homo-bilayers to independently control the vertical electric field ($F_w$) and the total doping density ($n$) in $WSe_2$ bilayers by the top and bottom gates. Both natural and twisted bilayers have been investigated. A natural bilayer is a stack of two monolayers with a 180° relative rotation [17-20]. And a twisted bilayer is a stack of two monolayers that are generally rotationally misaligned and do not have atomic registry [23]. Figure 1b shows a schematic of the device structure. Optical micrographs of two samples are shown in Supplementary Fig. S2. These devices have been built using a dry mechanical transfer method from individual van der Waals layers exfoliated from bulk crystals [24,25]. Bilayer $WSe_2$, electrically grounded, has been fully encapsulated for high sample quality between two hexagonal boron nitride (hBN) substrates of nearly identical thickness $d$, which also serve as the gate dielectrics. Graphene has been used as both contact and gate electrodes. The vertical field, or equivalently, the interlayer potential difference in $WSe_2$ bilayers ($\phi = -eF_w t$ with $e$ and $t$ denoting the elementary charge and layer separation, respectively), is controlled by the difference $\Delta V = V_{bg} - V_{tg}$ between the bottom gate voltage $V_{bg}$ and top gate voltage $V_{tg}$. On the other hand, the total doping density is governed by the sum of the two gate voltages. For more details see Supplementary Sect. 1.

In contrast to monolayer $WSe_2$, which is a direct band gap semiconductor [26,27], bilayer $WSe_2$ is an indirect gap semiconductor due to the finite interlayer coupling [17,26-29]. Figure 1c illustrates the schematic electronic band structure of natural bilayer $WSe_2$. The valence band maximum and the conduction band minimum are located at the K(K') and Q(Q') points of the Brillouin zone [29], respectively. The lowest-energy optical transitions are therefore indirect transitions between the K and Q valleys that require scattering with phonons or defects to conserve momentum [26, 27]. On the other hand, the lowest-energy direct optical transitions occur between the bands at the K(K') valleys, identical to those in monolayers. Similarly, twisted bilayers are also indirect gap semiconductors with the indirect gap energy dependent on the twist angle and atomic registry [23]. Figure 1e shows the reflectance contrast and photoluminescence (PL) spectra of a monolayer, natural bilayer and twisted bilayer $WSe_2$ device under zero vertical field. Whereas the reflectance is dominated by exciton resonance X corresponding to direct transitions at the K(K') valleys for all three samples, the PL spectrum of bilayers shows a lower-energy resonance $X_I$ (boxed) corresponding to indirect transitions that is absent in monolayer $WSe_2$ [28]. The emission features between X and $X_I$ arise from localized excitons, which are significant only at low temperatures. These observations are consistent with the band diagram of Fig. 1c, which has been further verified by systematic doping dependence studies of the reflectance spectra (See Supplementary Sect. 2.2). We note that the $X_I$ emission consists of two peaks, each of a width ~ 10 meV at low temperature and separated by ~ 17 meV. The separation energy is nearly independent of doping density, vertical field, optical excitation fluence, and sample temperature (below and Supplementary Sect. 2.3, 2.4 and 2.5), and matches well the distribution of the optical phonon density of states of $WSe_2$ [30]. We thus assign the high-energy peak of $X_I$ momentum-indirect exciton emission, and the low-energy peak, its phonon replica. Further systematic studies on the $X_I$ fine structures are warranted.

We investigate the effect of a vertical electric field on the exciton resonance energy and its recombination dynamics in bilayer $WSe_2$ by PL spectroscopy and time-resolved PL spectroscopy, respectively. Below we present the results for a natural bilayer device at 10 K. Qualitatively identical results have been observed in other devices of natural and twisted bilayers and the results for a twisted bilayer device are included in Supplementary Sect. 2.8. Figure 2a is a representative gate dependence of the PL spectra in a contour plot at a fixed electron doping density of $n = 1.3 \times 10^{12}$ cm$^{-2}$. The major features are that the $X_I$ emission redshifts linearly with the gate voltage difference $\Delta V$ by about 70 meV, and the effect is symmetric for the two opposing vertical directions. On the other hand, the energy of the intralayer exciton X shows a weak dependence on $\Delta V$ (Supplementary Fig. S4), in agreement with earlier reports [21, 31]. This comparison suggests that $X_I$ is a layer-polarized (i.e. spatially-indirect) exciton, as in coupled semiconductor quantum wells [32]. In the absence of a field or a potential difference between the layers, the $WSe_2$ bilayer electronic bands are layer degenerate, and the electron-hole (e-h) wave functions are a superposition of the layer eigenstates. However, under a finite potential difference the layer degeneracy is lifted (Fig. 1d) and the e-h wave functions are layer-polarized [17-19] (Fig. 1a). The polarity can be controlled by the direction of the applied field. The emission corresponding to the lowest-energy e-h recombination thus redshifts by an amount given by the absolute value of the interlayer potential difference, regardless of the field direction. Since the e-h wave functions are confined in the atomically thin $WSe_2$ samples, we refer to the effect quantum-confined Stark effect as in semiconductor quantum wells [33]. The major $X_I$ feature has been studied systematically as a function of doping density (Fig. 2b, positive $n$ for electron doping and negative $n$ for hole doping). Near the charge neutrality point, the Stark effect emerges immediately with $\Delta V$, whereas for a finite doping density, an appreciable effect is observed only for $\Delta V$ beyond a threshold value $\Delta V_0$. Our results show that $\Delta V_0$ generally increases with doping density and the exciton energy redshift rate is nearly doping independent (symbols, Fig. 2c).

The observed strong Stark effect and the existence of a threshold gate voltage difference can be understood in the following picture. In the presence of doping, free carriers are equally distributed between the two layers with the interlayer potential difference $\phi = 0$ if $\Delta V = 0$. When $\Delta V$ increases, more and more carriers are shifted into one of the layers, but $\phi$ is still small due to the large density of states or quantum capacitance of the constituent monolayers in bilayer WSe$_2$. Until $\Delta V$ reaches the threshold value $\Delta V_0$, at which all free carriers are located in one layer, the interlayer potential difference starts to rise rapidly. The threshold value increases with doping density since a larger gate voltage difference is required to fully layer polarize the free carriers. We express the interlayer potential difference based on a simple parallel-plate capacitor model (Supplementary Sect. 1 and Fig. 1b) for $\Delta V > 0$ as

$$\frac{\phi}{e} = \begin{cases} -\frac{C_g}{C_q + 2C_w + C_g}\Delta V, & \Delta V < \Delta V_0 \\ \frac{C_g}{2C_w + C_g}\left(\frac{C_q}{C_q + 2C_w + C_g}\Delta V_0 - \Delta V\right), & \Delta V \geq \Delta V_0 \end{cases} \quad (1)$$

where $\Delta V_0 = \frac{C_q + 2C_w + C_g}{C_q C_g}|n|e$. The result is symmetric about $\Delta V = 0$. Here $C_g = \frac{\varepsilon_{bn}\varepsilon_0}{d}$ and $C_w = \frac{\varepsilon_w \varepsilon_0}{t}$ ($\varepsilon_{bn}$ and $\varepsilon_w$ denote the out-of-plane dielectric constant of hBN and WSe$_2$, respectively, and $\varepsilon_0$ is the vacuum permittivity) are the geometric capacitance of the gates and bilayer WSe$_2$, respectively; and $C_q = \frac{2m^*e^2}{\pi \hbar^2}$ ($m^*$ and $\hbar$ denote the band mass and the Planck's constant, respectively) is the quantum capacitance of each WSe$_2$ monolayer. These capacitances can be evaluated from the measured hBN thickness $d \approx 17$ nm and the reported materials parameters ($\varepsilon_{bn} \approx 2.5$ [25, 31], $\varepsilon_w \approx 7.2$ [34], $t \approx 0.65$ nm [35] and $m^* \approx 0.5 m_0$ [29]) to be $C_q \approx 0.67 \times 10^{-4}$ Fcm$^{-2}$, $C_w \approx 0.98 \times 10^{-5}$ Fcm$^{-2}$ and $C_g \approx 1.32 \times 10^{-7}$ Fcm$^{-2}$. Equipped with these values we compare Eqn. (1) with experiment in Fig. 2b and 2c (dashed lines). The exciton energy in the absence of a field in Fig. 2b is the only free parameter. Despite its simplicity, the model captures the main features of the experiment remarkably well. More sophisticated models including the realistic band structure and Coulomb interaction effects are, however, required for a more quantitative description of the experimental result.

Next we examine the influence of the vertical field on the layer-polarized exciton dynamics. Figure 3a and 3b show the time-resolved PL of layer-polarized excitons (spectrally integrated) following a femtosecond pulse excitation at $\tau = 0$. Results for several gate voltage differences under an electron doping density of $3 \times 10^{11}$ cm$^{-2}$ (Fig. 3a) and a hole doping density of $3.4 \times 10^{12}$ cm$^{-2}$ (Fig. 3b) are included. In both examples, the exciton recombination lifetime $\tau_{PL}$ increases dramatically with $\Delta V$. In particular, for the hole doping case $\tau_{PL}$ exceeds the time interval between successive excitation pulses (12.5 ns) that is evident from the residual PL signal at $\tau < 0$. The recombination lifetime $\tau_{PL}$ has been extracted assuming exponential decays (see Method) and summarized in Fig. 3c and 3d. The lifetime increases by an order of magnitude (from 0.1 to 0.9 ns) for the electron-doping example and by more than two orders of magnitude (from 0.2 to 22 ns) for the hole-doping example. (These lifetimes are also significantly larger than 1-10's ps reported for monolayer WSe$_2$ [5, 6]). The large enhancement of the PL lifetime is correlated with the strong Stark effect shown in Fig. 2. Although a quantitative description of these results requires a detailed knowledge of the microscopic mechanisms for exciton nonradiative recombination (responsible for the observed PL decay) and is beyond the scope of this study, the general trend of the gate dependence of the exciton lifetime is expected for layer-polarized excitons. For $\Delta V > \Delta V_0$, the electron and hole wave functions become layer polarized [17] (Fig. 1a). The reduced spatial overlap of the *e-h* wave functions leads to longer lifetimes, as in coupled semiconductor quantum wells [32]. Note that the observed *e-h* asymmetry is likely related to the different nature of the donor and acceptor levels.

Finally, we explore the possibility of creating high-density exciton gases enabled by the long exciton recombination lifetimes in bilayer WSe$_2$ by continuous-wave optical pumping. Figure 4a shows the spectra of the X$_I$ emission under $\Delta V$ = 15.1 V. The emission energy continuously blueshifts with increasing pump power from 2 to 64 μW (note that heating alone produces a redshift as shown in Supplementary Sect. 2.4). A maximum blueshift of $\delta E \approx 2$ meV is observed under a pump intensity of ~ 6×10$^3$ Wcm$^{-2}$. Such a blueshift is a result of a repulsive dipole-dipole exciton interaction [36, 37] since the layer-polarized excitons have a permanent dipole along the vertical direction (Fig. 1a), opposing the externally applied field. It can be used to evaluate the exciton density $n_X$ [37]: $\delta E = \frac{n_X e^2}{C_w}$. By using the capacitance value from above, we obtain an average layer-polarized exciton density of $1.2 \times 10^{11}$ cm$^{-2}$ (corresponding to an exciton degeneracy temperature of ~ 5 K [22]). We note that $n_X$ evaluated using the dipole-dipole repulsion energy provides a lower bound on the exciton density because the exciton-exciton exchange and correlation energies have been neglected [36]. But what has limited the exciton density under optical pumping in this regime? In Fig. 4b we show the exciton blueshift $\delta E$, as well as the calibrated $n_X$, as a function of pump power under different gate voltage differences ($\Delta V \geq \Delta V_0$). We have limited the pump power such that laser heating is not significant. Under the same pump power, $n_X$ increases with $\Delta V$. This is consistent with the observed longer exciton lifetimes (Fig. 3). On the other hand, for any given $\Delta V$ the pump power dependence of $n_X$ turns from linear to sublinear. This suggests that a density-dependent recombination process becomes effective to limit the equilibrium exciton density. A plausible mechanism is the Auger recombination [38], which finds support in the power dependence of the time-resolved PL dynamics (Supplementary Sect. 2.9). Although the strong Auger effect may be inevitable in atomically thin samples [38], reducing the interlayer coupling by inserting a monolayer of large band gap semiconductors (e.g. WS$_2$ or hBN) between the WSe$_2$ monolayers could be employed to weaken the Auger process while maintaining appreciable *e-h* interactions to achieve higher exciton densities in future studies. Our results may have implications for the search of degenerate exciton gases [22] and new optoelectronics applications in 2D semiconductors [5, 6, 13, 14].

**Method**

**Photoluminescence and time-resolved photoluminescence spectroscopy.** Optical spectroscopy of dual-gated field-effect WSe$_2$ homo-bilayer devices in an optical cryostat was performed with a home-built confocal setup. An Olympus objective (N.A. = 0.6) was used to focus the excitation beam onto the devices and also collect the reflection/emission from them to achieve a sub-micron spatial resolution. For the reflectance contrast measurements, broadband radiation from a supercontinuum light source was employed with an incident power < 10 μW on the devices. The collected reflection was detected by a spectrometer equipped with a charge-coupled-device (CCD) camera. For the photoluminescence (PL) measurements, the excitation light source was a He-Ne laser at 633 nm with an incident power < 50 μW on the devices. A long-pass filter removed the laser line for detection of the PL spectrum by the spectrometer and CCD camera. The pump power on the devices was varied from 2 to 70 μW for the optical pumping experiment. For the time-resolved PL measurements, a modelocked Ti:sapphire oscillator centered at 729 nm with a repetition rate of 80 MHz and pulse duration of 50 fs (Spectra Physics) was employed. The collected PL was sent through a tunable longpass filter to eliminate the reflected laser beam and detected by the time-correlated single photon counting method with an avalanche photodiode and a PicoHarp system (PicoQuant). An average power of 1 μW on the devices was used to minimize the Auger process while maintaining a good signal-to-noise ratio. The instrument response function was measured by detecting the laser pulse using the same setup.

**Data analysis.** The PL spectra of layer-polarized excitons were fitted using two Lorentzian functions to account for the main feature and its phonon replica. The shift of the center of the high-energy feature (the major indirect exciton emission) was presented in this study. The shift of the low-energy feature (phonon replica) is nearly identical. Near the charge neutrality point, four emission peaks were present due to the presence of charge inhomogeneities, and four Lorentzian functions were used to fit the data (More details are provided in Supplementary Sect. 2.7). The time-resolved PL traces under electron doping were fitted using a single exponential function $Ae^{-\tau/\tau_{PL}}$ convoluted with the instrument response function. Here amplitude $A$ and PL lifetime $\tau_{PL}$ are the two free fitting parameters. The time-resolved PL traces under hole doping show a clear fast initial decay followed by a long decay for most $\Delta V$'s. A double exponential function was therefore employed to describe them. For small $\Delta V$'s, a single exponential decay was adequate to describe the dynamics. Figure 3d shows the lifetime or the lifetime of the long decay component (if a double exponential function was used). A double exponential analysis of the entire set of results with both lifetimes and the amplitude ratio is included in Supplementary Sect. 2.9. Details of the initial fast decay are also presented in Supplementary Sect. 2.9. Also, cyclic excitation with a known pulse separation (12.5 ns) was used to account for PL lifetimes that are longer than or comparable to the separation between two consecutive pulses.

**Figures**

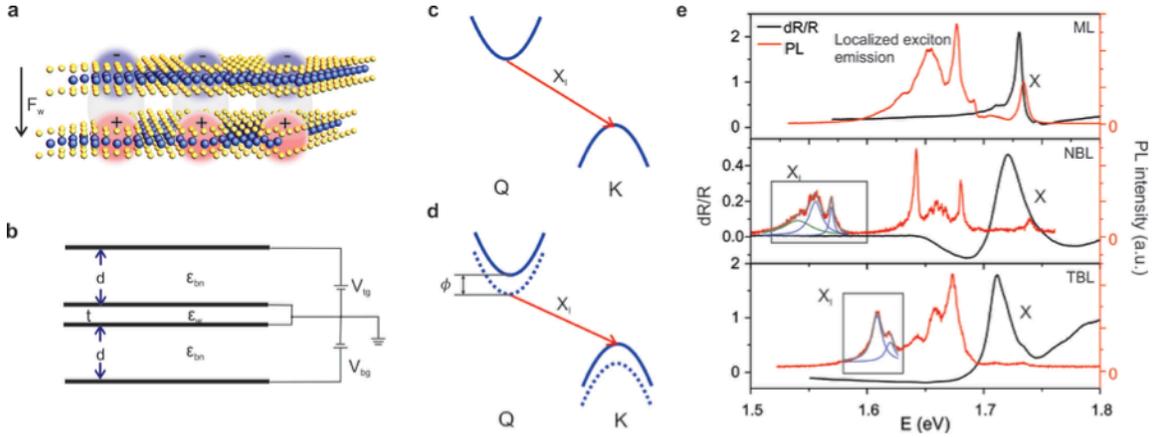

**Figure 1. Layer-polarized excitons in WSe$_2$ homo-bilayers. a,** Creation of layer-polarized excitons with the electron and hole wave functions separated into two different layers by an externally applied vertical electric field $F_W$. **b,** Schematic of dual-gated field-effect devices employed for optical studies. Natural or twisted WSe$_2$ bilayers separated by distance $t$ are grounded by a graphene contact and encapsulated by two hexagonal boron nitride (hBN) substrates of identical thickness $d$. Two graphene gates apply a top gate voltage $V_{tg}$ and bottom gate voltage $V_{bg}$. And $\varepsilon_{bn}$ and $\varepsilon_w$ are the out-of-plane dielectric constant of hBN and WSe$_2$, respectively. **c,** Schematic of the electronic band structure of pristine homo-bilayers with an indirect band gap. The valence band maximum and the conduction band minimum are located at the K(K') and the Q(Q') point of the Brillouin zone, respectively. The bands are layer degenerate. **d,** Under a finite gate voltage difference, the layer degeneracy of the bands is lifted by the interlayer potential difference with the solid and dotted lines denoting the bands of the top and bottom layers, respectively. The transitions shown in **c** and **d**, modified by the electron-hole interactions, correspond to the observed indirect exciton $X_I$ emission. **e,** Reflectance contrast (black lines) and photoluminescence spectra (red lines) of a representative monolayer (ML), nature bilayer (NBL) and twisted bilayer (TBL) WSe$_2$ device under zero field. Emission features corresponding to indirect exciton $X_I$ (boxed), localized excitons and exciton X are shown in order of increasing energy.

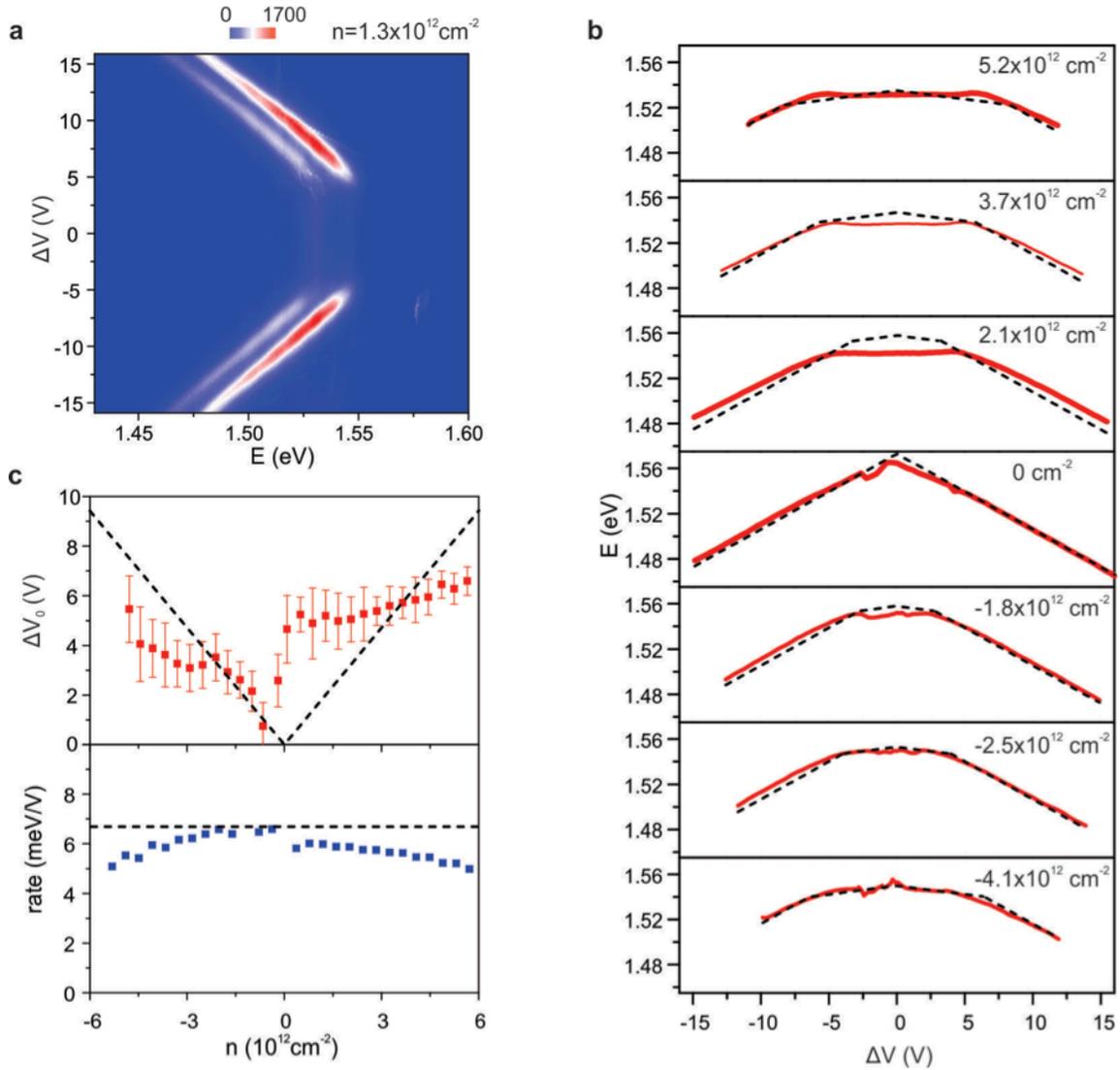

**Figure 2. Quantum-confined Stark effect. a,** Contour plot of the layer-polarized exciton emission intensity as a function of photon energy (bottom axis) and gate voltage difference $\Delta V$ (left axis). The total electron density in the bilayer is fixed at $1.3 \times 10^{12}$ cm$^{-2}$. **b,** Layer-polarized exciton emission energy as a function of gate voltage difference $\Delta V$ for differing doping densities. **c,** Threshold gate voltage difference $\Delta V_0$, beyond which the large Stark effect is observed, and the corresponding redshift rate of the layer-polarized exciton energy as a function of doping density. The dashed lines in **b** and **c** are predictions of Eqn. (1).

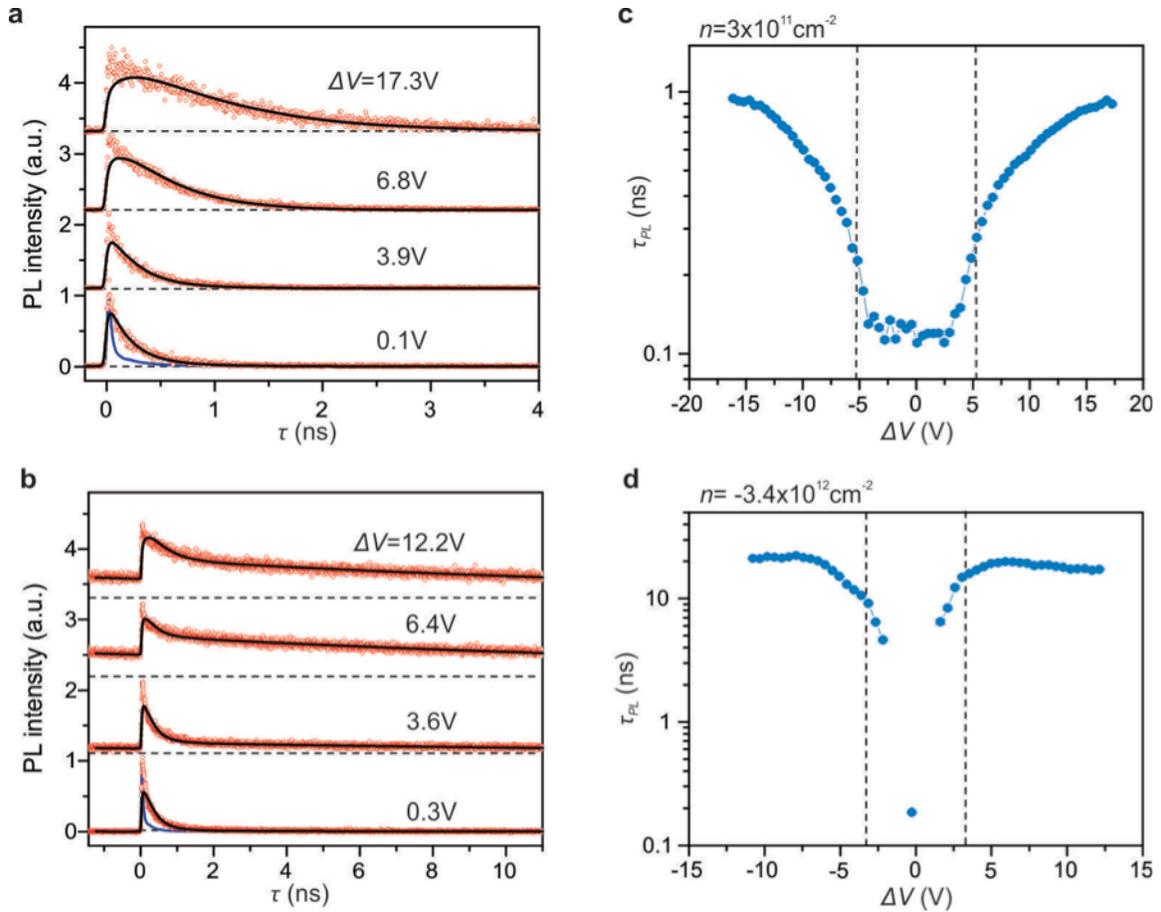

**Figure 3. Photoluminescence decay dynamics of layer-polarized excitons. a, b,** Time evolution of layer-polarized exciton emission following excitation by a femtosecond laser pulse at time $\tau = 0$ for a fixed electron density $3\times10^{11}\,\mathrm{cm^{-2}}$ (**a**) and hole density $3.4\times10^{12}\,\mathrm{cm^{-2}}$ (**b**). The traces for different gate voltage differences are vertically displaced for clarity. The black dashed lines indicate the baseline for each trace. The blue solid line is the instrument response. The red symbols are experimental data and the black solid lines are fits by exponential decay functions convoluted by the instrument response. **c, d,** The extracted PL lifetime $\tau_{PL}$ as a function of gate voltage difference. The error bars are the uncertainty associated with the least squares fitting. The dashed lines indicate the threshold voltage $\Delta V_0$ obtained from the Stark effect measurement of Fig. 2c.

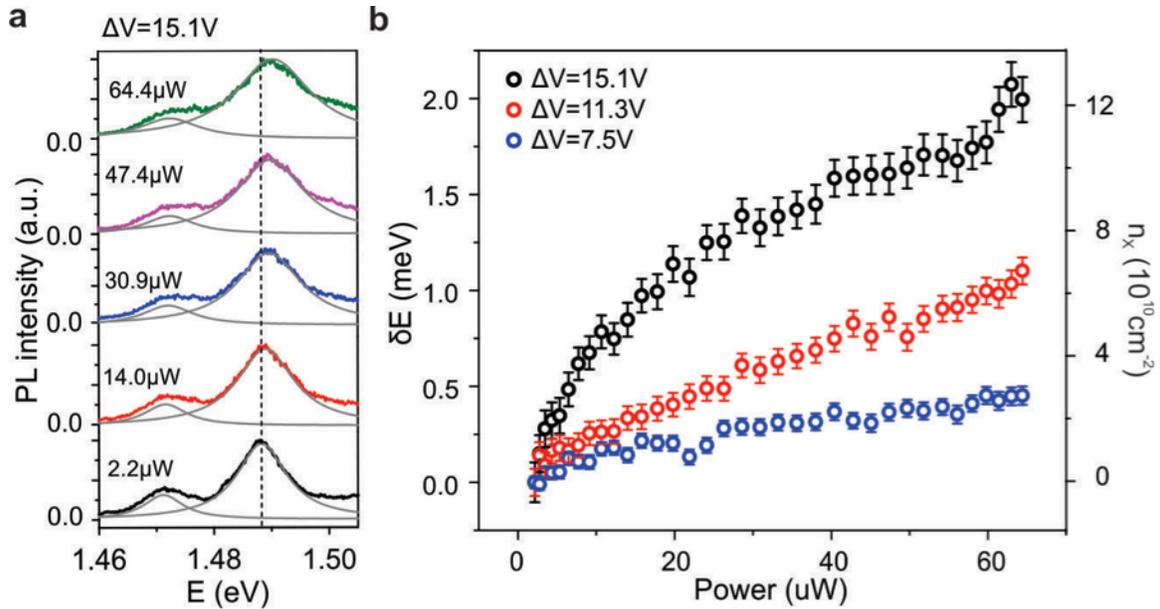

**Figure 4. Creation of layer-polarized exciton gases by continuous-wave optical pumping. a,** Evolution with pump power of the PL spectra of layer-polarized excitons under gate voltage difference $\Delta V$ = 15.1 V. The solid lines are double Lorentzian fits. A blueshift of the emission energy up to 2 meV is observed. **b,** Blueshift of the layer-polarized exciton energy (left), as well as the estimated exciton density (right), as a function of pump power for several representative gate voltage differences. Error bars are the uncertainty of the corresponding Lorentzian fitting parameter.

**References**

1. Geim, A. K.; Grigorieva, I. V. *Nature* **2013,** 499, (7459), 419-425.
2. Novoselov, K. S.; Mishchenko, A.; Carvalho, A.; Castro Neto, A. H. *Science* **2016,** 353, 6298.
3. Yu, H. Y.; Wang, Y.; Tong, Q. J.; Xu, X. D.; Yao, W. *Physical Review Letters* **2015,** 115, (18), 187002.
4. Li, Y.-M.; Li, J.; Shi, L.-K.; Zhang, D.; Yang, W.; Chang, K. *Physical Review Letters* **2015,** 115, (16), 166804.
5. Mak, K. F.; Shan, J. *Nat Photon* **2016,** 10, (4), 216-226.
6. Schaibley, J. R.; Yu, H.; Clark, G.; Rivera, P.; Ross, J. S.; Seyler, K. L.; Yao, W.; Xu, X. *Nature Review Materials* **2016,** 1, 16055.
7. Rivera, P.; Seyler, K. L.; Yu, H.; Schaibley, J. R.; Yan, J.; Mandrus, D. G.; Yao, W.; Xu, X. *Science* **2016,** 351, (6274), 688.
8. Wu, F.; Lovorn, T.; MacDonald, A. H. *Physical Review Letters* **2017,** 118, (14), 147401.
9. Hong, X.; Kim, J.; Shi, S.-F.; Zhang, Y.; Jin, C.; Sun, Y.; Tongay, S.; Wu, J.; Zhang, Y.; Wang, F. *Nat Nano* **2014,** 9, (9), 682-686.
10. Ceballos, F.; Ju, M.-G.; Lane, S. D.; Zeng, X. C.; Zhao, H. *Nano Letters* **2017,** 17, (3), 1623-1628.

11. Rivera, P.; Schaibley, J. R.; Jones, A. M.; Ross, J. S.; Wu, S.; Aivazian, G.; Klement, P.; Seyler, K.; Clark, G.; Ghimire, N. J.; Yan, J.; Mandrus, D. G.; Yao, W.; Xu, X. *Nature Communications* **2015,** 6, 6242.
12. Kim, J.; Jin, C.; Chen, B.; Cai, H.; Zhao, T.; Lee, P.; Kahn, S.; Watanabe, K.; Taniguchi, T.; Tongay, S.; Crommie, M. F.; Wang, F. *arXiv:1612.05359* **2016**.
13. Lee, C.-H.; Lee, G.-H.; van der Zande, A. M.; Chen, W.; Li, Y.; Han, M.; Cui, X.; Arefe, G.; Nuckolls, C.; Heinz, T. F.; Guo, J.; Hone, J.; Kim, P. *Nat Nano* **2014,** 9, (9), 676-681.
14. Withers, F.; Del Pozo-Zamudio, O.; Mishchenko, A.; Rooney, A. P.; Gholinia, A.; Watanabe, K.; Taniguchi, T.; Haigh, S. J.; Geim, A. K.; Tartakovskii, A. I.; Novoselov, K. S. *Nat Mater* **2015,** 14, (3), 301-306.
15. Wilson, N. R.; Nguyen, P. V.; Seyler, K.; Rivera, P.; Marsden, A. J.; Laker, Z. P. L.; Constantinescu, G. C.; Kandyba, V.; Barinov, A.; Hine, N. D. M.; Xu, X.; Cobden, D. H. *Science Advances* **2017,** 3, (2), e1601832.
16. Zhang, C.; Chuu, C.-P.; Ren, X.; Li, M.-Y.; Li, L.-J.; Jin, C.; Chou, M.-Y.; Shih, C.-K. *Science Advances* **2017,** 3, (1), e1601459.
17. Ramasubramaniam, A.; Naveh, D.; Towe, E. *Physical Review B* **2011,** 84, (20), 205325.
18. Wu, S.; Ross, J. S.; Liu, G.-B.; Aivazian, G.; Jones, A.; Fei, Z.; Zhu, W.; Xiao, D.; Yao, W.; Cobden, D.; Xu, X. *Nat Phys* **2013,** 9, (3), 149-153.
19. Jones, A. M.; Yu, H.; Ross, J. S.; Klement, P.; Ghimire, N. J.; Yan, J.; Mandrus, D. G.; Yao, W.; Xu, X. *Nat Phys* **2014,** 10, (2), 130-134.
20. Lee, J.; Mak, K. F.; Shan, J. *Nat Nano* **2016,** 11, (5), 421-425.
21. Klein, J.; Wierzbowski, J.; Regler, A.; Becker, J.; Heimbach, F.; Müller, K.; Kaniber, M.; Finley, J. J. *Nano Letters* **2016,** 16, (3), 1554-1559.
22. Fogler, M. M.; Butov, L. V.; Novoselov, K. S. *Nature Communications* **2014,** 5, 4555.
23. van der Zande, A. M.; Kunstmann, J.; Chernikov, A.; Chenet, D. A.; You, Y.; Zhang, X.; Huang, P. Y.; Berkelbach, T. C.; Wang, L.; Zhang, F.; Hybertsen, M. S.; Muller, D. A.; Reichman, D. R.; Heinz, T. F.; Hone, J. C. *Nano Letters* **2014,** 14, (7), 3869-3875.
24. Cui, X.; Lee, G.-H.; Kim, Y. D.; Arefe, G.; Huang, P. Y.; Lee, C.-H.; Chenet, D. A.; Zhang, X.; Wang, L.; Ye, F.; Pizzocchero, F.; Jessen, B. S.; Watanabe, K.; Taniguchi, T.; Muller, D. A.; Low, T.; Kim, P.; Hone, J. *Nat Nano* **2015,** 10, (6), 534-540.
25. Wang, Z.; Shan, J.; Mak, K. F. *Nat Nano* **2017,** 12, (2), 144-149.
26. Splendiani, A.; Sun, L.; Zhang, Y.; Li, T.; Kim, J.; Chim, C.-Y.; Galli, G.; Wang, F. *Nano Letters* **2010,** 10, (4), 1271-1275.
27. Mak, K. F.; Lee, C.; Hone, J.; Shan, J.; Heinz, T. F. *Physical Review Letters* **2010,** 105, (13), 136805.
28. Zhao, W.; Ghorannevis, Z.; Chu, L.; Toh, M.; Kloc, C.; Tan, P.-H.; Eda, G. *ACS Nano* **2013,** 7, (1), 791-797.
29. Liu, G.-B.; Xiao, D.; Yao, Y.; Xu, X.; Yao, W. *Chemical Society Reviews* **2015,** 44, (9), 2643-2663.
30. Amin, B.; Kaloni, T. P.; Schwingenschlogl, U. *RSC Advances* **2014,** 4, (65), 34561-34565.
31. Wang, Z.; Zhao, L.; Mak, K. F.; Shan, J. *Nano Letters* **2017,** 17, (2), 740-746.

32. Lai, C.-W. E. *Lawrence Berkeley National Laboratory Thesis* **2004**.
33. Miller, D. A. B.; Chemla, D. S.; Damen, T. C.; Gossard, A. C.; Wiegmann, W.; Wood, T. H.; Burrus, C. A. *Physical Review Letters* **1984,** 53, (22), 2173-2176.
34. Kim, K.; Larentis, S.; Fallahazad, B.; Lee, K.; Xue, J.; Dillen, D. C.; Corbet, C. M.; Tutuc, E. *ACS Nano* **2015,** 9, (4), 4527-4532.
35. Bromley, R. A.; Murray, R. B.; Yoffe, A. D. *Journal of Physics C: Solid State Physics* **1972,** 5, (7), 759.
36. Zimmermann, R.; Schindler, C. *Solid State Communications* **2007,** 144, (9), 395-398.
37. Stern, M.; Garmider, V.; Umansky, V.; Bar-Joseph, I. *Physical Review Letters* **2008,** 100, (25), 256402.
38. Sun, D.; Rao, Y.; Reider, G. A.; Chen, G.; You, Y.; Brézin, L.; Harutyunyan, A. R.; Heinz, T. F. *Nano Letters* **2014,** 14, (10), 5625-5629.